\documentstyle[11pt]{article}
\begin{document}
\title{Decouple a Coupled KdV System of Nutku and O\~{g}uz}

\author{Heng Chun Hu and Q.P. Liu\\[0.1cm]
\small{China University of Mining and Technology,}\\
\small{100083, Beijing,
 China}\thanks{mailing address}\\
\small{and}\\
\small{International Centre for Theoretical Physics},\\
\small{Strada Costiera 11,
34014 Trieste, Italy}}
\date{}
\maketitle{}
\begin{abstract}
A coupled  KdV system with a free parameter proposed by Nutku
and O\~{g}uz is considered. It is shown that the system passes
the WTC's Painlev\'e test for arbitrary value of the parameter.
A further analysis yields that the parameter can be scaled away
and the system can be decoupled.
\end{abstract}

Soliton or integrable equations are those systems which have rich structures,
 such as infinite number of conservation laws, multi-Hamiltonian structures,
  infinite symmetries, B\"acklund transformations. An important question is
   how integrable systems couple without loss of integrability. The first
   coupled KdV system was proposed by Hirota and Satsuma in 1981 \cite{hs}.
   Since then, many coupled KdV systems have been constructed. We here
   mention Ito's system \cite{ito}, the coupled systems resulted from
   Drinfeld-Sokolov framework \cite{ds}, those from the group theoretical
   approach of Kyoto group \cite{jm} and the ones constructed by means of
   energy dependent Schr\"odinger operator \cite{af}.

Nutku and O\~{g}uz \cite{no} considered the following system
\begin{equation}\label{no1}
q_t=q_{xxx}+2aqq_x+rr_{x}+(qr)_x,\quad r_t=r_{xxx}+2brr_x+qq_x+(qr)_x,
\end{equation}
where $a$ and $b$ are constants. They pointed out that this
system (\ref{no1}) decouples if $a=\mp b$. It is further shown that subject
to
\begin{equation}\label{con}
a+b=1,
\end{equation}
the system (\ref{no1}) is a bi-Hamiltonian system with two local Hamiltonian
structures. (We notice that the dispersionless version of this system with
the condition (\ref{con}) was studied recently by Matsuno \cite{matsuno}.)
Thus, one has a bi-Hamiltonian system which contains a free parameter.
Our motivation is to consider the integrability of this system. In general,
a bi-Hamiltonian system is supposed to be integrable since it has an infinite
 number of conserved quantities. However, it is peculiar that a system is
 integrable with arbitrary value of parameter. We believe that either the
 system is integrable for a certain value of the parameter or the parameter
 can be removed. For the system of Nutku and O\~{g}uz we will show that it
 is the latter case. WTC's Painlev\'e analysis will be used to explore the
 system. (See \cite{conte} and the references there for Painlev\'e property
 and applications.)

We first convert the system (\ref{no1}) into a more convenient form.
Taking $q+r$ and $q-r$ as the new dependent variables and then rescaling
them, we find that the system (\ref{no1}) under the condition (\ref{con})
is equivalent to
\begin{equation}\label{no}
u_t=u_{xxx}+uu_x+vv_x,\quad v_t=v_{xxx}+\lambda vv_x+(uv)_x,
\end{equation}
where $\lambda$ is a parameter which is assumed as real. If $\lambda=0$,
the system (\ref{no}) is the complexly coupled KdV system discussed by
Fuchssteiner \cite{fu}. In this case, the system decouples.

Now we follow Weiss, Tabor and Carnevale \cite{wtc} and perform a Painlev\'e
analysis for (\ref{no}). By leading analysis: $u\to u_0\phi^\alpha$,
$v\to v_0\phi^{\beta}$, we obtain
\begin{eqnarray}
\alpha=-2,& &\beta=-2,\nonumber \\
u_0=\frac{-6(\delta\lambda+\sqrt{\lambda^2+4})}{\sqrt{\lambda^2+4}}\phi_{x}^2,&&
v_0=\frac{12\delta}{\sqrt{\lambda^2+4}}\phi_{x}^{2},
\label{uv0}
\end{eqnarray}
where $\delta^2=1$.

Thus we make the following expansions
\begin{equation}\label{exp}
u=\phi^{-2}\sum_{i=0}^{\infty}u_i\phi^{i}, \quad
v=\phi^{-2}\sum_{i=0}^{\infty}v_i\phi^{i},
\end{equation}
it is found that resonances are
\[
-1,2, 3, 4, 4, 6
\]
and we have only one branch. To show that this branch has the Painlev\'e
property, we use Kruskal's simplification, that is $\phi(x,t)=t+f(x)$. Then,
the conditions required at all resonances satisfy identically. Therefore,
we conclude that the system (\ref{no}) passes the WTC's Painlev\'e test for
arbitrary $\lambda$.

To gain more information on (\ref{no}), we define the transformations by
truncating the series expansions (\ref{exp}) on the constant level as follows
\begin{equation}\label{trun}
u={u_0\over \phi^2}+{u_1\over \phi}+u_2,\quad
v={v_0\over \phi^2}+{v_1\over \phi}+v_2.
\end{equation}
For definiteness, we will concentrate on the case $\delta=1$. Inserting
above expressions for $u$ and $v$ into the system (\ref{no}) and setting the
coefficients of each power of $\phi$ to zero, one obtains a system of
equations. The coefficients of $\phi^{-5}$ provide us the formulae for $u_0$
and $v_0$ (\ref{uv0}), while the coefficients of $\phi^{-4}$ and $\phi^0$
yield
\begin{equation}\label{uv1}
u_1=\frac{6(\lambda+\sqrt{\lambda^2+4})}{\sqrt{\lambda^2+4}}\phi_{xx},\quad
v_1=\frac{-12}{\sqrt{\lambda^2+4}}\phi_{xx},
\end{equation}
and
\begin{equation}
u_{2t}=u_{2xxx}+u_2u_{2x}+v_2v_{2x},\quad
v_{2t}=v_{2xxx}+\lambda v_2v_{2x}+(u_2v_2)_x.
\end{equation}
The rest of equations, after eliminating $u_0$, $v_0$ by (\ref{uv0}) and
$u_1$, $v_1$ by (\ref{uv1}), read
\begin{eqnarray}
[\phi_{xx}u_2+\frac{1}{2}(\lambda-\sqrt{\lambda^2+4})\phi_{xx}v_2-(\phi_t-
\phi_{xxx})_x]_x&=&0,\label{9}\\
\phi_x(\phi_xu_{2x}+3u_2\phi_{xx})+\frac{1}{2}(\lambda-
\sqrt{\lambda^2+4})\phi_x(\phi_xv_{2x}+3v_2\phi_{xx})+&&\nonumber\\
+5\phi_x\phi_{xxxx}-2\phi_x\phi_{xt}-2\phi_{xx}\phi_{xxx}-\phi_t\phi_{xx}&=&0,\\
\phi^{2}_{x}u_2+\frac{1}{2}(\lambda-\sqrt{\lambda^2+4})\phi_{x}^{2}v_2-
\phi_t\phi_{x}+4\phi_x\phi_{xxx}-3\phi_{xx}^2 &=&0.\label{11}
\end{eqnarray}
We observe that in above three equations (\ref{9}-\ref{11}),
the fields $u_2$ and $v_2$ can be combined into a single variable:
$u_2+\frac{1}{2}(\lambda-\sqrt{\lambda^2+4})v_2$. This observation
suggests a transformation of variables as follows
\begin{equation}
\tilde{Q}=u+\frac{1}{2}(\lambda-\sqrt{\lambda^2+4})v,
\quad \tilde{R}=v,
\end{equation}
then it is easy to verify that the following equation holds
\begin{equation}
\tilde{Q}_t=\tilde{Q}_{xxx}+\tilde{Q}\tilde{Q}_x, \quad
\tilde{R}_t=\tilde{R}_{xxx}+\sqrt{\lambda^2+4}\tilde{R}\tilde{R}_x
+(\tilde{Q}\tilde{R})_x.
\end{equation}
A simple rescaling of $\tilde{Q}$ and $\tilde{R}$
\[
\tilde{Q}=Q, \quad \tilde{R}={R\over\sqrt{\lambda^2+4}},
\]
leads to the following system
\begin{equation}
Q_t=Q_{xxx}+QQ_x, \quad R_t=R_{xxx}+RR_x+(QR)_x
\end{equation}
a system free from any parameter. We now notice that the first equation is
nothing but the celebrated KdV equation while the second one is the KdV
equation with an extra term. Now it is easy to see that by introducing
\[
Q=S, \quad Q+R=T
\]
both $S$ and $T$ solve the KdV equation.

Thus, by means of the WTC's Painlev\'e analysis, we find that coupled KdV
system (\ref{no}) is equivalent to two KdV equations. In this way,
the original parameter is removed and the system is decoupled. Therefore,
we have a better understanding of the structure of (\ref{no}).
This example further shows that the WTC's approach is a powerful method
to study nonlinear systems.

\bigskip

{\bf Acknowledgements}

One of us (QPL) discussed the Nutku-O\~{g}uz system with Andrew Pickering
a few years ago. He would like to thank both Andrew Pickering and Senyue Lou
for interesting conversions. We also should like to thank International
Centre for Theoretical Physics for hospitality.
This work was also partially supported
by the Ministry of Education and the National Natural Science
Foundation of China under grant number 19971094.


\begin{thebibliography}{99}
\bibitem{af} M. Antonowicz and A. P. Fordy, Physica D {\bf 28} (1986) 345;
Commun. Math. Phys. {\bf 124} (1989) 465.
\bibitem{conte} R. Conte (ed.), {\em The Painlev\'e Property:
One Century Later} (Springer, 1999).
\bibitem{ds} V.G. Drinfeld and V.V. Sokolov, Soviet J.  Math. {\bf 30} (1985) 1975.
\bibitem{fu}  B. Fuchssteiner, Prog. Theor. Phys. {\bf 68} (1982) 1082.
\bibitem{hs} R. Hirota and J. Satsuma, Phys. Lett. {\bf 85A} (1981) 407.
\bibitem{ito} M. Ito, Phys. Lett. {\bf 91A} (1982) 335.
\bibitem{jm} M. Jimbo and T. Miwa, Publ. RIMS, Kyoto Univ. {\bf 19} (1983) 943.
\bibitem{matsuno} Y. Matsuno,  J. Math. Phys. {\bf 42} (2001) 1744.
\bibitem{no} Y. Nutku and \"{O}. O\~{g}uz, Il Nuovo Cimento {\bf 105B} (1990) 1381.
\bibitem{wtc} J. Weiss, M. Tabor and G. Carnevale,  J. Math. Phys. {\bf 24} (1984) 522.
\end{thebibliography}
\end{document}